# Active analog tuning of the phase of light in the visible regime by bismuth-based metamaterials

Marina García-Pardo,[1] Eva Nieto Pinero,[1] Amanda K. Petford-Long,[2,3] Rosalía Serna,[1,4] Johann Toudert[1,5]

[1]*Laser Processing Group, Instituto de Óptica, IO, CSIC, Madrid, Spain*

[2]*Materials Science Division, Argonne National Laboratory, Argonne, IL, USA*

[3]*Materials Science and Engineering Dept, Northwestern Univ, Evanston, IL, USA*

[4]*rosalia.serna@csic.es*

[5]*johann.toudert@gmail.com*

**Abstract.** Active and analog tuning of the phase of light is needed to boost the switching performance of photonic devices. However, demonstrations of this type of tuning in the pivotal visible spectral region are still scarce. Herein we report active analog tuning of the phase of visible light reflected by a bismuth-based metamaterial, enabled by a reversible solid-liquid transition. This metamaterial, fabricated by a lithography-free approach, consists of two-dimensional assemblies of polydisperse plasmonic bismuth nanostructures embedded in a refractory and transparent aluminum oxide matrix. Analog tuning of the phase is achieved by controlled heating of the metamaterial to melt a fraction of the nanostructures. A maximum tuning of 320º (1.8π) is observed upon complete melting of the nanostructures at 230ºC. This tuning is reversible by cooling to 25ºC. In addition, it presents a wide hysteretic character due to liquid bismuth undercooling. This enables the phase achieved by this analog approach to remain stable over a broad temperature range upon cooling and until re-solidification occurs around 100ºC. Therefore, bismuth-based metamaterials are appealing for applications including optical data storage with enhanced information density or bistable photonic switching with a tunable "on" state.

**Introduction.** Metamaterials enabling *active* tuning of the amplitude and phase of light are needed to boost the performance of photonic devices such as optical modulators, reconfigurable antennas, polarization controllers, displays, and optical data storage media. [1] In this context, phase change metamaterials (PCMs) based on compounds such as $VO_2$ or $GeSb_xTe_y$ (GSTs) have been thoroughly considered, because their optical properties can be tuned *reversibly* via *externally triggered phase transitions*. [2-19] The optical properties of $VO_2$ change upon its monoclinic/tetragonal transition. The tetragonal phase forms upon heating above 65ºC and the monoclinic phase recovers upon cooling below this temperature. [2-7] This "*volatile*" character is an asset for optical modulators, polarization controllers, or reconfigurable antennas. However, for applications such as optical data storage, GSTs are preferred because of their "*non-volatile*" amorphous/crystalline transition. In contrast with $VO_2$, the crystalline phase formed upon heating above 250ºC remains upon cooling. [8-19] However, returning to the original amorphous phase requires energetic processes such as melt-quenching near 600ºC or irradiation with electrons or light pulses. [8,10,12,15,16,18]

Interestingly, although tuning based on a phase transition conveys a *binary* picture involving an "off" and an "on" state, multilevel and even *analog* tuning of the optical properties of PCMs were recently reported. [4,6,7,13,14,16-18] This opens the way to photonic devices with outstanding features such as color that can be tuned actively and in an analog manner, or optical data storage with enhanced information density. The fine tuning reported in these works was made possible by triggering the phase transition of a controlled volume fraction of the material, to achieve a controlled proportion of both phases. This in turn enabled the analog tuning of the amplitude of light transmitted by Mie and anapole resonant GST metasurfaces, [17] and of the phase of light reflected by $Au/VO_2/Au$ gap plasmon metasurfaces. [6] Such tuning was demonstrated in the mid and near infrared, respectively. In fact, most of the reports showing PCMs with actively tunable optical properties focus on these spectral regions, where the optical response of $VO_2$ and GSTs is the most strongly affected by their crystallinity. [2-4,6,8,9,11-15,17] Meanwhile, demonstrations of PCMs enabling an active analog tuning of the amplitude *and* phase of light in the pivotal *visible* and *ultraviolet* regions are still scarce. Only recently have some reports showed active multilevel or analog color tuning with PCMs based on $VO_2$ and GST nanolayers. [7,16,18]

As an alternative to $VO_2$ and GSTs, in previous works we have proposed to use bismuth (Bi) nanostructures as the active building blocks of PCMs enabling the tuning of ultraviolet and visible light. [20-22] Bi is an outstanding optical material that presents ultraviolet-visible plasmonic properties induced by giant interband transitions in the solid state. [21, 23-26] Upon heating above 270ºC, it undergoes a solid/liquid transition that rubs out these interband transitions, so that liquid Bi behaves as a lossy Drude metal. [27] The solid/liquid transition thus markedly changes the spectrum of the Bi dielectric function in the ultraviolet and visible, as shown in **Fig. 1a**. This change induces a shift in the plasmon resonances of the Bi nanostructures, which translates into a significant change in the ultraviolet-visible optical properties of the PCM. When the nanostructures melt, their size, shape and organization remain unchanged, because in the considered PCMs the Bi nanostructures are embedded in a refractory matrix that acts as a solid mold. This matrix also enables the nanostructures to return to their initial solid state upon decreasing the temperature. [20, 28] Based on this behavior, in a previous work we reported the applicability of Bi-based PCMs for high-contrast binary switching of the amplitude of light. [22] However, the potential of Bi-based PCMs for analog tuning and phase tuning remained to be explored.

Herein we demonstrate the active analog tuning of the phase of visible light reflected by a Bi-based PCM. Such tuning is achieved by controlled heating of the PCM at a selected temperature to melt a fraction of the embedded Bi nanostructures. A maximum tuning of the phase of light of 320º (1.8π) is observed upon complete melting of the nanostructures at 230ºC. The tuning is fully reversible by cooling to 25ºC, yet with a wide hysteretic response due to Bi undercooling, which is interesting for applications such as optical data storage with enhanced information density or for bistable photonic switching with a tunable "on" state.

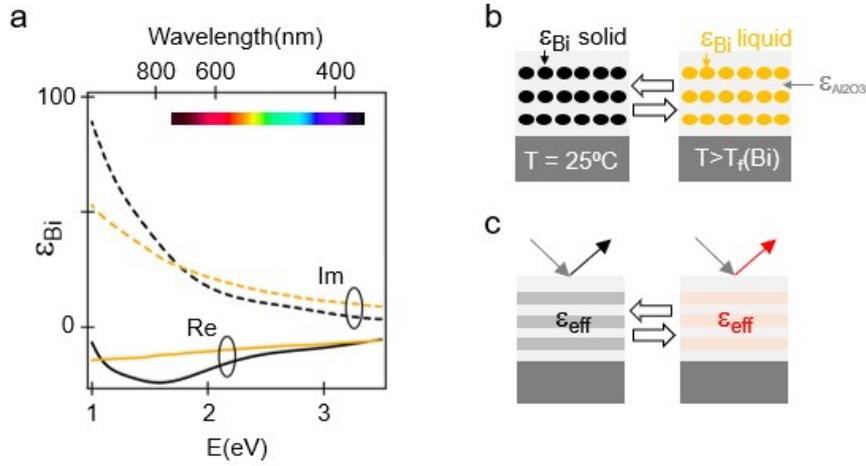

**Figure 1. Working principle of the Bi-based PCM.** (a) Spectra of the real and imaginary part of the dielectric function $\varepsilon_{Bi}$ of solid Bi (black lines) and liquid Bi (orange lines). (b) Simplified cross-section drawing of the PCM design consisting of a multilayer stack alternating composite $Bi:Al_2O_3$ nanolayers with spacer layers made of $Al_2O_3$. Each composite nanolayer consists of a two-dimensional assembly of Bi nanostructures embedded in $Al_2O_3$. Upon increasing temperature above the Bi melting point, the nanostructures melt and their dielectric function changes from that of solid Bi to that of liquid Bi. (c) Therefore, upon melting the Bi nanostructures, the effective dielectric function $\varepsilon_{eff}$ of the composite nanolayers changes and so do the optical reflection properties of the PCM. Note that the dielectric function $\varepsilon_{Al2O3}$ of $Al_2O_3$ does not change significantly over the range of temperatures considered.

The PCM was designed according to the simplified cross-section drawing in **Fig. 1b**. It consists of a multilayer stack of alternating $Bi:Al_2O_3$ composite nanolayers and $Al_2O_3$ spacer layers, deposited on a reflective substrate (Si). Each composite nanolayer consists of a two-dimensional assembly of Bi nanostructures embedded in $Al_2O_3$. For this experiment we have selected $Al_2O_3$ as embedding medium because of its excellent thermal stability and transparency, however other refractory and transparent materials could be used. As shown in **Fig. 1c**, upon Bi melting the effective dielectric function of the composite nanolayers changes significantly. This results in a change in the optical reflection properties of the whole PCM, which are driven by the interplay between plasmon resonances in the composite nanolayers and optical interference in the multilayer stack. [29] This multilayer configuration enables fabrication of metamaterials with remarkable optical properties with no need for costly or low-throughput lithography techniques.

**PCM structure and temperature-dependent optical properties.** Bi-based PCMs were produced by pulsed laser deposition, a fully lithography-free material fabrication technique. A cross-section transmission electron microscopy image of one of the fabricated Bi-based PCMs is shown in **Fig. 2a**. The light-contrast areas in the image correspond to the $Al_2O_3$ spacer layers. The $Al_2O_3$ is amorphous. The 5 dark-contrast areas correspond to the $Bi:Al_2O_3$ composite nanolayers, which present an approximate thickness of 6 nm. Each of these nanolayers consists of a two-dimensional assembly of Bi nanostructures embedded in $Al_2O_3$. The nanostructures show slightly oblate ellipsoidal shapes with a polydisperse size distribution.

This PCM was placed on the temperature-controlled stage of a spectroscopic ellipsometer as shown in **Fig. 2b**, to measure its optical properties as a function of temperature T (25-240ºC), angle of incidence AOI (20º-75º) and photon energy E (1.0-3.5 eV) covering the visible and part of the near infrared wavelength range (350-1240 nm). Measurements were first carried out under static conditions, i.e. after stabilizing T at a fixed value.

**Fig. 2c** shows the spectra of the effective dielectric function of the Bi:Al$_2$O$_3$ composite nanolayers at 25ºC and 230ºC. The significant contrast seen between these spectra results from the different state of the Bi nanostructures at T = 25ºC and 230ºC, where they are solid and liquid, respectively. The Bi nanostructures are fully melted at 230ºC, i.e. 40ºC below the bulk Bi melting point. This agrees with the size-dependence of the Bi solid/liquid transition reported in previous works. [30]

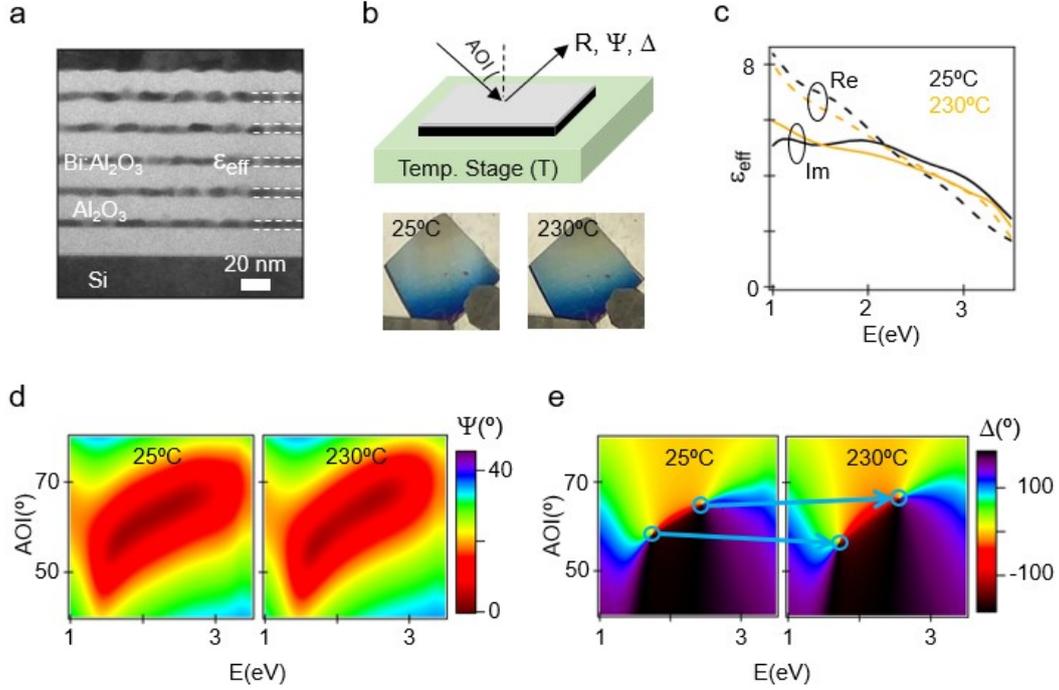

**Figure 2. Structure and temperature-dependent optical properties of the fabricated Bi-based PCM.** (a) Cross-section image of the fabricated PCM obtained by transmission electron microscopy. The composite nanolayers consist of polydisperse Bi nanostructures embedded in Al$_2$O$_3$. (b) Configuration used for temperature-dependent optical measurements, and macroscopic images of the PCM taken at 25ºC and 230ºC. (c) Spectra of the effective dielectric function of the nanocomposite layers at 25ºC and 230ºC. (d) Interpolated maps of the ellipsometric amplitude coefficient, Ψ, of the PCM as a function of photon energy E and angle of incidence AOI, at 25ºC and 230ºC. (e) Corresponding maps of the ellipsometric phase coefficient, Δ. The maps show two singularities (marked with circles) that are displaced upon increasing temperature (arrows), as a result of the Bi nanostructures melting. Near these singularities, the phase of light reflected by the PCM varies strongly with temperature.

Although it does not strongly affect the visual aspect of the PCM surface (see image **Fig. 2b**), the melting of the Bi nanostructures significantly affects the ellipsometric response of the PCM, as shown in **Figs. 2d** and **Fig. 2e**. **Fig. 2d** displays interpolated maps of the ellipsometric amplitude coefficient Ψ versus E and AOI, at T = 25ºC and 230ºC. Similar features are seen at both temperatures, in agreement with the small change in visual aspect, with very small Ψ values in a broad range of E and for AOIs between 55 and 65º. The interpolated maps of the ellipsometric phase coefficient Δ shown in **Fig. 2e** correlate with those of Ψ. However, the maps of Δ display two singularities located at (i: 2.45 eV, 64.5º) and (ii: 1.75 eV, 58.5º) when T = 25ºC. Within the energy range of these singularities, there are strong vortex-like variations of Δ in the E–AOI plane, as large as 360º (phase change of 2π). The singularities are markedly displaced when T is increased to 230ºC: they are at (i: 2.55 eV, 66.5º) and at (ii: 1.72 eV, 56.5º). This implies that, at E and AOI values close to these singularities, the phase of light reflected by the PCM varies strongly when the temperature is increased, and the Bi nanostructures melt. This effect is closely related to the concept of topological optical darkness. [31-33] Herein, we take advantage of this effect to demonstrate a broad tuning of the phase of light by temperature control.

**Broad phase tuning by temperature control.** To analyze and quantify this broad tuning, we studied the optical properties of the PCM at several temperatures from 25ºC to 230ºC, for an AOI of 64.5º. This AOI was chosen to approach the singularity (i) predicted by the interpolated Δ map of **Fig. 2d**. Measurements were done in static conditions at fixed T's as previously. **Fig. 3a** displays the measured Ψ spectra, which all show a clear minimum when E is close to 2.35 eV. Such minimum is due to the near-cancelation of the p-polarized reflectance ($R_p$) of the PCM when E is close to 2.35 eV, which is seen in **Fig. 3b**. This near-cancelation of $R_p$ implies that the phase $\delta_p$ of p-polarized reflected light undergoes a sharp jump in the spectrum for photon energies near 2.35 eV. [34–36] In contrast, the phase $\delta_S$ of s-polarized reflected light is nearly photon energy-independent. Therefore, the spectrum of $\Delta = \delta_P - \delta_S$ shown in **Fig. 3c** displays a sharp jump when E is close to 2.35 eV, which is due to the jump in $\delta_P$. As seen in **Fig. 3c**, upon increasing T, the Δ jump reverses (and so does the sign of Δ) and then damps. At photon energies close to the jump, Δ is very sensitive to T and the variation of Δ with T comes exclusively from that of $\delta_P$. Therefore, our results show that, at photon energies close to 2.35 eV, the phase of p-polarized light reflected by the PCM can be broadly tuned by controlling temperature.

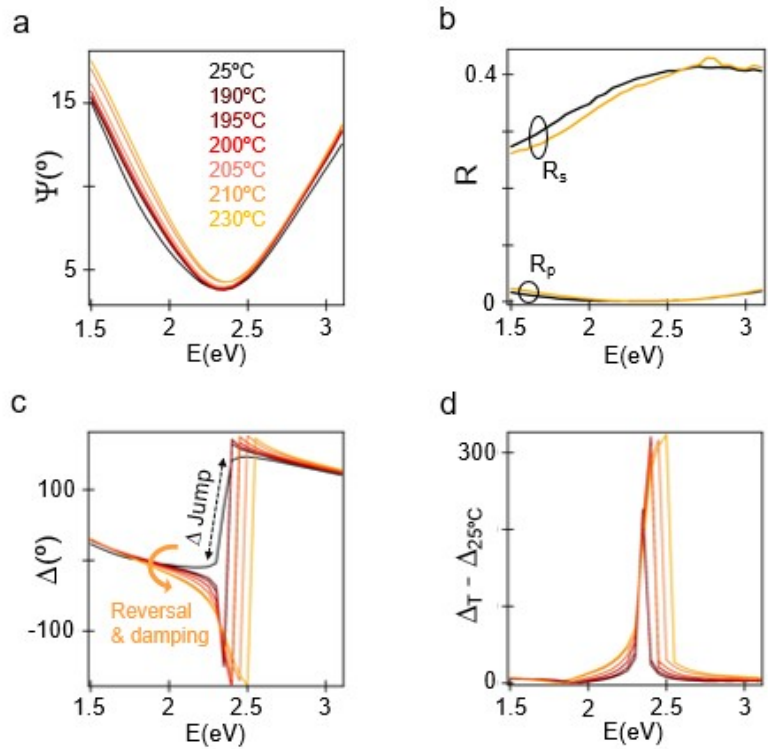

**Figure 3. Broad and analog tuning of the phase of visible p-polarized light reflected by the Bi-based PCM.** (a) Ψ spectra of the PCM for an AOI of 64.5º measured at different temperatures between 25ºC and 230ºC. (b) Corresponding p-polarized and s-polarized reflectance spectra (at 25ºC and 230ºC only). Near 2.35 eV, $R_P$ nearly cancels ($R_P$ = 0.17% and 0.14% at 25ºC and 230ºC) (c) Corresponding Δ spectra, which show a jump near 2.35 eV. The initial jump at room temperature is marked by a double arrow. Upon increasing temperature, the jump reverses (and so does the sign of Δ) and damps. This yields a continuous variation of Δ with temperature that is due to the progressive melting of the Bi nanostructures. (d) Corresponding Δ variation with respect to its value at 25ºC, showing a maximum of 320º at 230ºC, when the Bi nanostructures are fully melted. The Δ variation with temperature comes exclusively from a variation in the phase of p-polarized reflected light.

The tuning range of Δ (equivalent to that of $\delta_P$) is quantified in **Fig. 3d** with respect to the initial Δ value at 25ºC. Upon increasing T to 230ºC, a variation in Δ that is larger than 180º (π) is observed at photon energies between 2.35 and 2.5 eV. This spectral window, which corresponds to wavelengths between 527 and 496 nm, is broad enough to be accessible with cost-effective thermally-tunable green laser diodes. The maximum Δ variation reached is 320° (1.8π) and this variation occurs upon a T variation of a few °C between 210 and 230°C. The sensitivity of Δ to T is less marked in a range of photon energies detuned from the Δ jump. For convenience, we focus on this lower sensitivity range to provide evidence for the analog character of the phase tuning. This character is for instance confirmed at E = 2.3 eV, where Δ decreases continuously upon increasing T, yet with a saturation when T approaches 230°C.

**Analog tuning, reversibility and hysteretic behavior.** We propose that this analog phase tuning is made possible by the fact that the Bi nanostructures do not all melt at a single temperature. Since the Bi melting point is size-dependent [30] and the Bi nanostructures are polydisperse in size, heating the PCM to T < 230°C melts only a fraction of the nanostructures. As shown in **Fig. 4a**, the proportion of solid and liquid Bi nanostructures in the PCM is set by the heating temperature. This enables analog tuning of the optical properties of the PCM, and therefore of the phase $\delta_P$ of p-polarized reflected light represented here by the ellipsometric phase coefficient Δ.

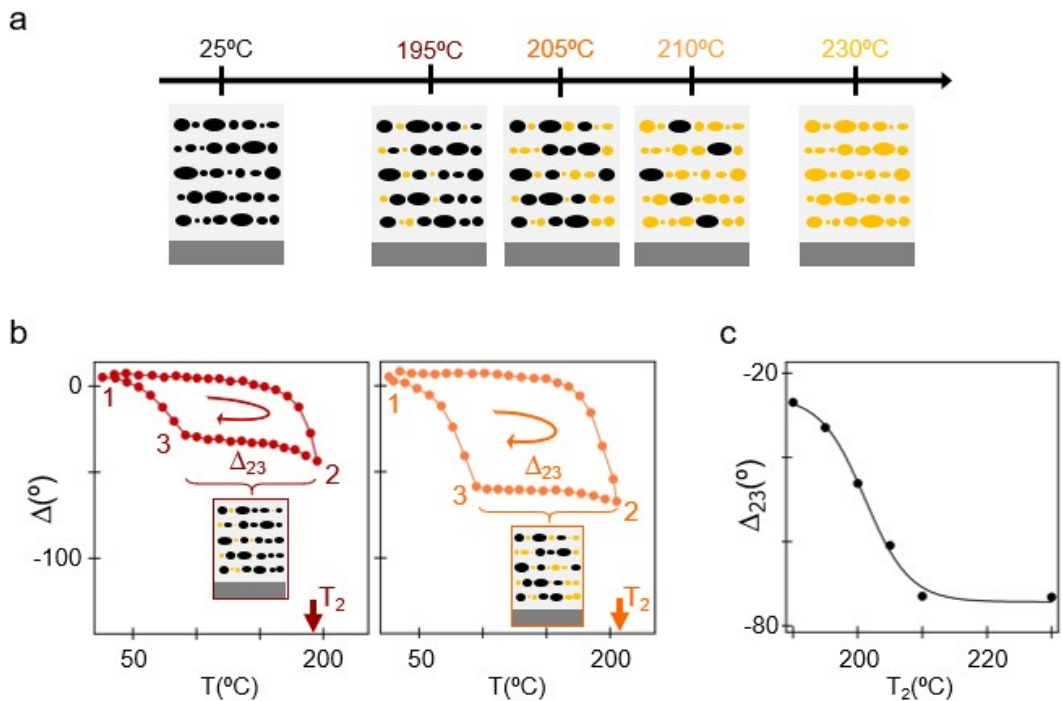

Figure 4. Analog, reversible and hysteretic character of the phase tuning. (a) Proposed microscopic origin for the analog character of the phase tuning. The Bi melting point is size-dependent, and so a controlled fraction of the polydisperse Bi nanostructures can be melted by heating the PCM to a given temperature. At 25°C, 100% of the nanostructures are solid (black). At 230°C, 100% are liquid (orange). (b) Δ = f(T) curves for the PCM recorded continuously when heating to a temperature $T_2$ and then cooling to 25°C. Left panel: $T_2$ = 195°C; Right panel: $T_2$ = 205°C. These curves show the reversibility and hysteretic character of the phase tuning. The Bi nanostructures melted at point 2 remain liquid when decreasing the temperature to point 3 (undercooling). This leaves Δ unchanged at the value $\Delta_{23}$ over a broad temperature range (~100°C). (c) Such a stable value of $\Delta_{23}$ can be tuned in an analog manner by controlling the fraction of melted nanostructures, i.e. by properly choosing $T_2$. All the data in this figure were recorded for E = 2.3 eV and AOI = 64.5°.

To ascertain the analog character of the phase tuning, and to show its reversibility and hysteretic properties, we finally measured Δ at a single value of E (2.3 eV) and AOI (64.5°) as a function of T under dynamic conditions. In this case Δ was continuously monitored while heating from room temperature $T_1$ = 25°C to a selected maximum value $T_2$ < 230°C and then while cooling down to room temperature $T_1$. The left panel of **Fig. 4b** displays the Δ = f(T) curve recorded for $T_2$ = 195°C. Upon heating, Δ starts decreasing when T ~ 185°C, i.e. when the smallest Bi nanostructures melt. By increasing T further to a selected value, a proportion of the Bi nanostructures melt so that $\Delta$ is decreased to a finely tuned value. This shows the fully analog character of the phase tuning. After reaching $T_2$, cooling down leaves $\Delta$ unchanged at a value $\Delta_{23}$ until T reaches $T_3 \sim 90°C$. We also carried out control experiments to check that this effect is not due to thermal inertia. In these experiments, cooling was stopped to stabilize T between $T_2$ and $T_3$. In such static conditions, which were maintained during more than 1 hour, $\Delta$ remained unchanged at $\Delta_{23}$. Therefore, we attribute the persistence of the $\Delta$ value to undercooling, which was shown to occur for Bi nanostructures. [28] Finally, upon further cooling to $T_1$, $\Delta$ returns to its initial value. The $\Delta = f(T)$ curve thus demonstrates the reversibility of the phase tuning, together with its wide hysteretic behavior ($\delta T \sim 100°C$).

Furthermore, we found it particularly interesting that the $\Delta_{23}$ value (and thus the height of the hysteresis) is set by the maximum heating temperature $T_2$. This can be seen by comparing the left and right panels of **Fig. 4b**, which show the $\Delta = f(T)$ curves for $T_2 = 195°C$ and $205°C$, respectively. Increasing $T_2$ enables melting of a larger fraction of Bi nanostructures. Correspondingly, this leaves a larger fraction of undercooled Bi nanostructures upon cooling down and translates into an increased hysteresis height. In line with the previously demonstrated analog phase tuning properties, we find that $\Delta_{23}$ can be finely tuned as seen in **Fig. 4c**. Therefore, the phase of the p-polarized reflected light can be tuned to the chosen value by heating the PCM to a selected temperature, and the phase value obtained remains stable when cooling down over a broad temperature range of about 100°C.

**Conclusions.** In summary, we have reported that Bi-based PCMs enable active and analog tuning of the phase of p-polarized reflected light at photon energies in the visible. The PCM is formed by a multilayer structure consisting of two-dimensional assemblies of polydisperse Bi nanostructures embedded in a refractory and transparent $Al_2O_3$ matrix. This nanostructured metamaterial has been fully grown by pulsed laser deposition, without the need for lithography techniques. Based on this PCM, we have demonstrated analog phase tuning by heating to a suitably chosen temperature to melt some of the Bi nanostructures. A maximum phase tuning of 320° (1.8π) has been obtained at 230°C by fully melting all the Bi nanostructures. After cooling down to 25°C, the Bi nanostructures return to their initial solid state thus demonstrating the full reversibility of the PCM operation. We checked this reversibility during repeated experiments over a period of one year, thus demonstrating the stability of the system. In addition, upon decreasing the temperature, undercooling of the Bi nanostructures provides the PCM with a wide hysteretic response that leaves the analog-tuned phase stable over a temperature range of about 100°C.

Based on these findings, we propose that Bi-based PCMs are especially appealing candidates for applications requiring an analog and non-volatile active tuning of the amplitude and phase of visible light. Indeed, although we considered a metamaterial structure that enables a broad phase tuning, Bi-based metamaterials can also be designed to achieve a broad amplitude tuning. [21] Possible applications include devices such as optical data storage platforms with enhanced information density, or bistable photonic switches with tunable "on" state. For such applications, Bi-based PCMs are advantageous over GSTs because returning the Bi nanostructures to their initial solid state does not require any high temperature melt-quench process. An optimum energetic efficiency will be achieved by building the Bi-based PCMs from smaller Bi nanostructures that undergo solidification closer to room temperature.

**Methods.** The Bi-based PCMs were grown by alternate pulsed laser deposition from pure Bi and $Al_2O_3$ targets sputtered sequentially onto a Si substrate. To grow each two-dimensional assembly of Bi nanostructures and each intercalated $Al_2O_3$ layer, the Bi target and $Al_2O_3$ target were irradiated with 300 and 6000 laser pulses, respectively. Further details about the deposition setup and procedure are given in ref. 23. Cross-section images were obtained with a Tecnai F20 transmission electron microscope, after focused ion beam lift-out and milling a piece of the PCM to electron transparency.

The temperature-dependent optical properties of the PCM were measured with a Woollam VASE spectroscopic ellipsometer in a Polarizer/Retarder/Sample/Rotating Analizer configuration, equipped with an Instec temperature control stage. During experiments in static conditions, we checked that the actual temperature of the PCM measured with a thermocouple in contact with the sample surface is equal to the one measured by the temperature control stage sensor.

To determine the effective dielectric function spectra of the composite nanolayers shown in Fig. 2b, the spectra of the ellipsometric coefficients $\Psi$ and $\Delta$ were first measured for E between 1 and 3.5 eV with a 0.05 eV step, at AOIs from 25º to 75º with a 10º step. These measurements were done in static conditions at 25ºC and 230ºC. At each temperature, all the measured spectra were fitted simultaneously using a single multilayer model mimicking the real layered structure as seen in the cross-section images. The thicknesses of the composite and $Al_2O_3$ spacer layers were set to the values measured from these images. The spectra of the dielectric functions of $Al_2O_3$ and Si were previously determined at each temperature from ellipsometry measurements of an $Al_2O_3$ film grown on Si and of a bare Si substrate, respectively. The dielectric function of the nanocomposite layers was modeled as the sum of 11 Kramers-Kronig consistent Lorentz oscillators and a real offset. The amplitude, width and energy of these oscillators and the real offset were left free during the fit. These were the only free fit parameters. An excellent fit quality was obtained at both temperatures (MSE < 8). With the best fit values of the parameters, $\Psi$ and $\Delta$ were simulated as a function of E and AOI to draw the high-resolution maps shown in Fig. 2c. This approach is equivalent to interpolating the experimental data measured at discrete E and AOIs.

The $\Psi$ and $\Delta$ spectra shown in Fig. 3 were also measured in static conditions. The measurements at each temperature were done after heating the PCM from 25ºC at a 2ºC/min rate and waiting 10 min to ensure thermal stabilization. After measurement at each temperature the PCM was cooled down at a 2ºC/min rate. The $\Delta = f(T)$ curves shown in Fig. 4 were measured in dynamic conditions, i.e. $\Delta$ was continuously recorded while increasing and then decreasing temperature with a rate 2ºC/min rate.

**Acknowledgements**

This research was supported by Spanish grants RTI2018-096498-B-I00 (MCIU/AEI/FEDER, UE) and LINKA20044 (CSIC). AKPL is grateful to the National Science Foundation under Collaborative grant #DMR 1600837 for funding. Use of the Center for Nanoscale Materials, an Office of Science user facility, was supported by the U.S. Department of Energy, Office of Science, Office of Basic Energy Sciences, under Contract No. DE-AC02-06CH11357.